\documentclass[twocolumn]{article}
    
\usepackage{subfigure,amsmath,xspace}
\usepackage{booktabs,multirow,xspace}
\usepackage{bmpsize}
\usepackage{graphicx}
\usepackage{amsthm}
\usepackage{dsfont}        
\usepackage{exscale}       
\usepackage{amsmath}       
\usepackage{amssymb}
\usepackage{color} 
\usepackage{hyperref} 
\usepackage[all]{hypcap} 
\definecolor{darkred}{rgb}{0.5,0,0}
\definecolor{darkgreen}{rgb}{0,0.5,0}
\definecolor{darkblue}{rgb}{0,0,0.5}

\hypersetup{
linktocpage,
bookmarksnumbered=true,
colorlinks,
linkcolor=darkblue,
filecolor=darkgreen,
urlcolor=darkred,
citecolor=darkblue,
breaklinks=true,
pdfauthor={Shiqi Dong (shiqidong@bupt.edu.cn)},
pdftitle={Invisible Steganography via Generative Adversarial Networks},
pdfsubject={},
raiselinks,
}

\usepackage{booktabs}

\def\imagenet{ImageNet\xspace}
\def\lfw{LFW\xspace}
\def\pascal{PASCAL-VOC12\xspace}
\def\ssim{SSIM\xspace}
\def\mssim{MS-SSIM\xspace}
\def\mse{MSE\xspace}
\def\bpp{bpp\xspace}
\def\psnr{PSNR\xspace}
\def\isgan{ISGAN\xspace}

\def\secref#1{Sec.~\ref{sec:#1}}

\begin{document}   
%
%
\pagestyle{headings}  
%

%
\title{Invisible Steganography via Generative Adversarial Networks
\thanks{This work was supported by the National key Research and Development Program of China(No.2016YFB0800404) and the NSF of China(U1636112,U1636212).}
}



\author{Ru Zhang         \and
        Shiqi Dong \thanks{Shiqi Dong is the corresponding author.} \and Jianyi Liu\\\\
        School of Cyberspace Security, \\
		Beijing University of Posts and Telecommunications, China\\
		shiqidong@bupt.edu.cn\\
}

\date{}

\maketitle

\begin{abstract}
Nowadays, there are plenty of works introducing convolutional neural networks (CNNs) to the steganalysis and exceeding conventional steganalysis algorithms. These works have shown the improving potential of deep learning in information hiding domain. There are also several works based on deep learning to do image steganography, but these works still have problems in capacity, invisibility and security. In this paper, we propose a novel CNN architecture named as \isgan to conceal a secret gray image into a color cover image on the sender side and exactly extract the secret image out on the receiver side. There are three contributions in our work: (i) we improve the invisibility by hiding the secret image only in the Y channel of the cover image; (ii) We introduce the generative adversarial networks to strengthen the security by minimizing the divergence between the empirical probability distributions of stego images and natural images. (iii) In order to associate with the human visual system better, we construct a mixed loss function which is more appropriate for steganography to generate more realistic stego images and reveal out more better secret images. Experiment results show that \isgan can achieve start-of-art performances on \lfw, \pascal and \imagenet datasets.

\end{abstract}

\section{Introduction}\label{sec1}

Image steganography is the main content of information hiding. The sender conceal a secret message into a cover image, then get the container image called stego, and finish the secret message's transmission on the public channel by transferring the stego image. Then the receiver part of the transmission can reveal the secret message out. Steganalysis is an attack to the steganography algorithm. The listener on the public channel intercept the image and analyze whether the image contains secret information. Since their proposed, steganography and steganalysis promote each other's progress.

Image steganography can be used into the transmission of secret information, watermark, copyright certification and many other applications. In general, we can measure a steganography algorithm by capacity, invisibility and security. The capacity is measured by bits-per-pixel (bpp) which means the average number of bits concealed into each pixel of the cover image. With the capacity becomes larger, the security and the invisibility become worse. The invisibility is measured by the similarity of the stego image and its corresponding cover image. The invisibility becomes better as the similarity going higher. The security is measured by whether the stego image can be recognized out from natural images by steganalysis algorithms. Correspondingly, there are two focused challenges constraining the steganography performance. The amount of hidden message alters the quality of stego images. The more message in it, the easier the stego image can be checked out. Another keypoint is the cover image itself. Concealing message into noisy, rich semantic region of the cover image yields less detectable perturbations than hiding into smooth region. 

Nowadays, traditional steganography algorithms, such as S-UNIWARD~\cite{s-uniward}, J-UNIWARD~\cite{s-uniward}, conceal the secret information into cover images' spatial domain or transform domains by hand-crafted embedding algorithms successfully and get excellent invisibility and security. With the rise of deep learning in recent years, deep learning has become the hottest research method in computer vision and has been introduced into information hiding domain. Volkhonskiy et al.~\cite{SGAN} proposed a steganography enhancement algorithm based on GAN, they concealed secret message into generated images with conventional algorithms and enhanced the security. But their generated images are warping in semantic, which will be drawn attention easily. Tang et al.~\cite{ASDL-GAN} proposed an automatic steganographic distortion learning framework, their generator can find pixels which are suitable for embedding and conceal message into them, their discriminator is trained as a steganalyzer. With the adversarial training, the model can finish the steganography process. But this kind of method has low capacity and is less secure than conventional algorithms. Baluja~\cite{nips} proposed a convolutional neural network based on the structure of encoder-decoder. The encoder network can conceal a secret image into a same size cover image successfully and the decoder network can reveal out the secret image completely. This method is different from other deep learning based models and conventional steganography algorithms, it has large capacity and strong invisibility. But stego images generated by this model is distorted in color and its security is bad. Inspired by Baluja's work, we proposed an invisible steganography via generative adversarial network named \isgan. Our model can conceal a gray secret image into a color cover image with the same size, and our model has large capacity, strong invisibility and high security. Comparing with previous works, the main contributions of our work are as below:

1. In order to suppress the distortion of stego images, we select a new steganography position. We only embed and extract secret information in the Y channel of the cover image. The color information is all in Cr and Cb channels of the cover image and can be saved completely into stego images, so the invisibility is strengthened. 

2. From the aspect of mathematics, the difference between the empirical probability distributions of stego images and natural images can be measured by the divergence. So we introduce the generative adverasial networks to increase the security throughout minimizing the divergence. In addition, we introduce several architectures from classic computer vision tasks to fuse the cover image and the secret image together better and get faster training speed. 

3. In order to fit the human visual system (HVS) better, we introduce the structure similarity index (SSIM)~\cite{SSIM} and its variant to construct a mixed loss function. The mixed loss function helps to generate more realistic stego images and reveal out better secret images. This point is never considered by any previous deep-learning-based works in information hiding domain. 

The rest of the paper is organized as follows. \secref{2} discusses related works, \secref{3} introduces architecture details of \isgan and the mixed loss function. \secref{4} gives details of different datasets, parameter settings, our experiment processes and results. Finally, \secref{5} concludes the paper with relevant discussion. 

\section{Related Works}\label{sec:2}

\paragraph{Steganalysis} There have been plenty of works using deep learning to do image steganalysis and got excellent performance. Qian et al.~\cite{GNCNN} proposed a CNN-based steganalysis model GNCNN, the model introduced the hand-crafted KV filter to extract residual noise and used the gaussian activation function to get more useful features. The performance of the GNCNN is inferior to the state-of-the-art hand-crafted feature set spatial rich model (SRM)~\cite{SRM} slightly. Based on GNCNN, Xu et al.~\cite{XuNet} presented Batch Normalization~\cite{BN} in to prevent the network falling into the local minima. XuNet was equipped with Tanh, $1\times1$ convolution, global average pooling, and got comparable performance to SRM~\cite{SRM}. Ye et al.~\cite{YeNet} put forward YeNet which surpassed SRM and its several variants. YeNet used 30 hand-crafted filters from SRM to prepropose images, applied well-designed activation function named TLU and selection-channel module to strengthen features from rich texture region where is more suitable for hiding information. Zeng et al.~\cite{Zeng} proposed a JPEG steganalysis model with less parameters than XuNet and got better performance than XuNet. These works have applied deep learning to steganalysis successfully, but there is still space for improvement.

\paragraph{Steganography} Since its introduction, generative adversarial networks\cite{GAN} have received more and more attention, achieved the state-of-art performance on tasks such as image generation, style transfer, speech synthesis and so on. The earliest application of deep learning to steganography was based on GAN. Volkhonskiy et al.~\cite{SGAN} proposed a DCGAN-based~\cite{DCGAN} model SGAN. SGAN consists of a generator network for generating cover images, a discriminator network for discriminating generated images from real images and a steganalyzer network for steganalysis. Hiding information in cover images generated by SGAN is securer than in natural images. Shi et al.~\cite{SSGAN} proposed SSGAN based on WGAN~\cite{WGAN}, their work was similar to SGAN and got better outcome. However, stego images generated by models similar to SGAN and SSGAN are warping in semantic and are more easily to draw attention than natural images, although these models reduce the detection rate of steganalysis algorithms. Tang et al.~\cite{ASDL-GAN} proposed an automatic steganographic distortion learning framework named as ASDL-GAN. The generator can translate a cover image into an embedding change probability matrix and the discriminator incorporates the XuNet architecture. In order to fit the optimal embedding simulator as well as propagate the gradient in back propagation, they proposed a ternary embedding simulator (TES) activation function. ASDL-GAN can learn steganographic distortions automatically, but its performance is inferior to S-UNIWARD. Yang et al.~\cite{Yang} improved ASDL-GAN and achieved better performance than S-UNIWARD. They used Selection-Channel-Aware (SCA)~\cite{YeNet} in generator as well as the U-Net framework~\cite{unet} which is introduced from the medical images segmentation. However, ASDL-GAN still refers too many prior knowledge from conventional steganography algorithms and its capacity is small. Hayes~\cite{Hayes} proposed a GAN-based model to hide a secret message into a cover image, and could reveal the secret message by his decoder successfully, but the invisibility is weak. 

Baluja~\cite{nips} designed a CNN model to conceal a color secret image into a color cover image yielding state-of-art performance. Atique et al.~\cite{Atique} proposed another encoder-decoder based model to finish the same steganography task (their secret images are gray images). This is a novel steganography method which gets rid of hand-crafted algorithms. It can learn how to merge the cover image and the secret image together automatically. But stego images generated by their models are distorted in color. As shown in Fig.~\ref{fig:LFWshow}, Atique's stego images are yellowing when compared with the corresponding cover images. And their stego images are easily recognized by well trained CNN-based steganalyzer~\cite{nips} because of the large capacity. Inspired by works of Baluja and Atique, we improve each shortcoming and get \isgan.

\section{Our Approach}\label{sec:3}

\begin{figure*}[!ht]
\begin{center}
\includegraphics[width=\textwidth]{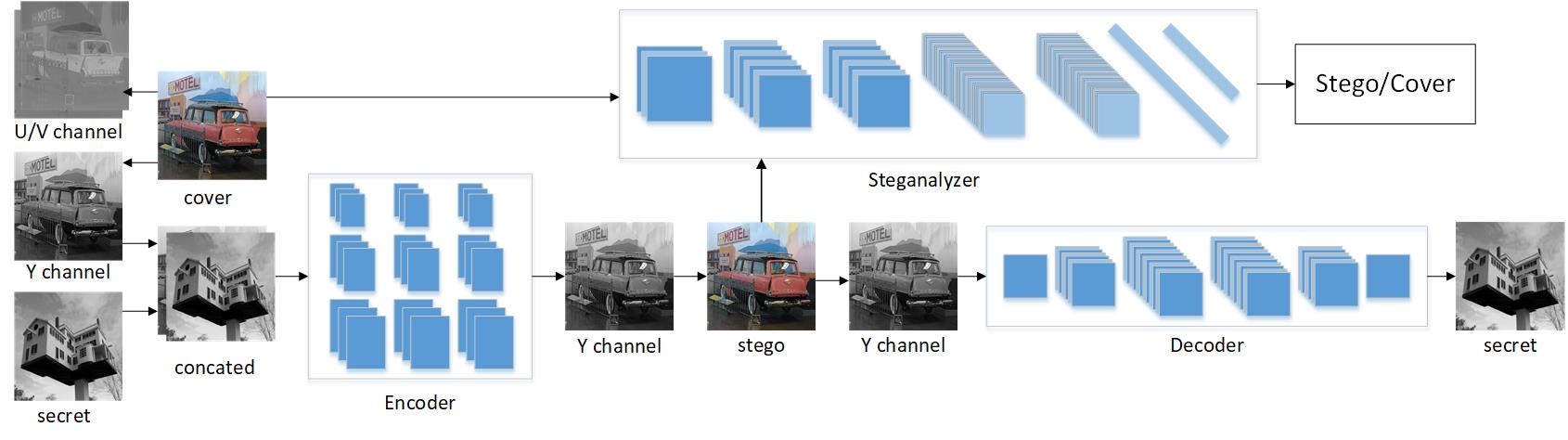}
 \caption{The overall architecture. The encoder network conceals a gray secret image into the Y channel of a same size cover image, then the Y channel output by the encoder net and the U/V channels constitute the stego image. The decoder network reveals the secret image from the Y channel of the stego image. The steganalyzer network tries to distinguish stego images from cover images thus improving the overall architecture's security.}
\label{fig:overall-architecture}  
\end{center}
\end{figure*}

The complete architecture of our model is shown in Fig.~\ref{fig:overall-architecture}. In this section, the new steganography position is introduced firstly. Then we discuss about our design considerations on the basic model and show specfic details of the encoder and the decoder. Thirdly, we present why the generative adversarial networks can improve the security and details of the discriminator. Finally, we explain the motivation to construct the mixed loss function. 

\subsection{New Steganography Position}\label{sec:3.1}

\begin{figure}[t]
  \centering
  \includegraphics[width=0.5\textwidth]{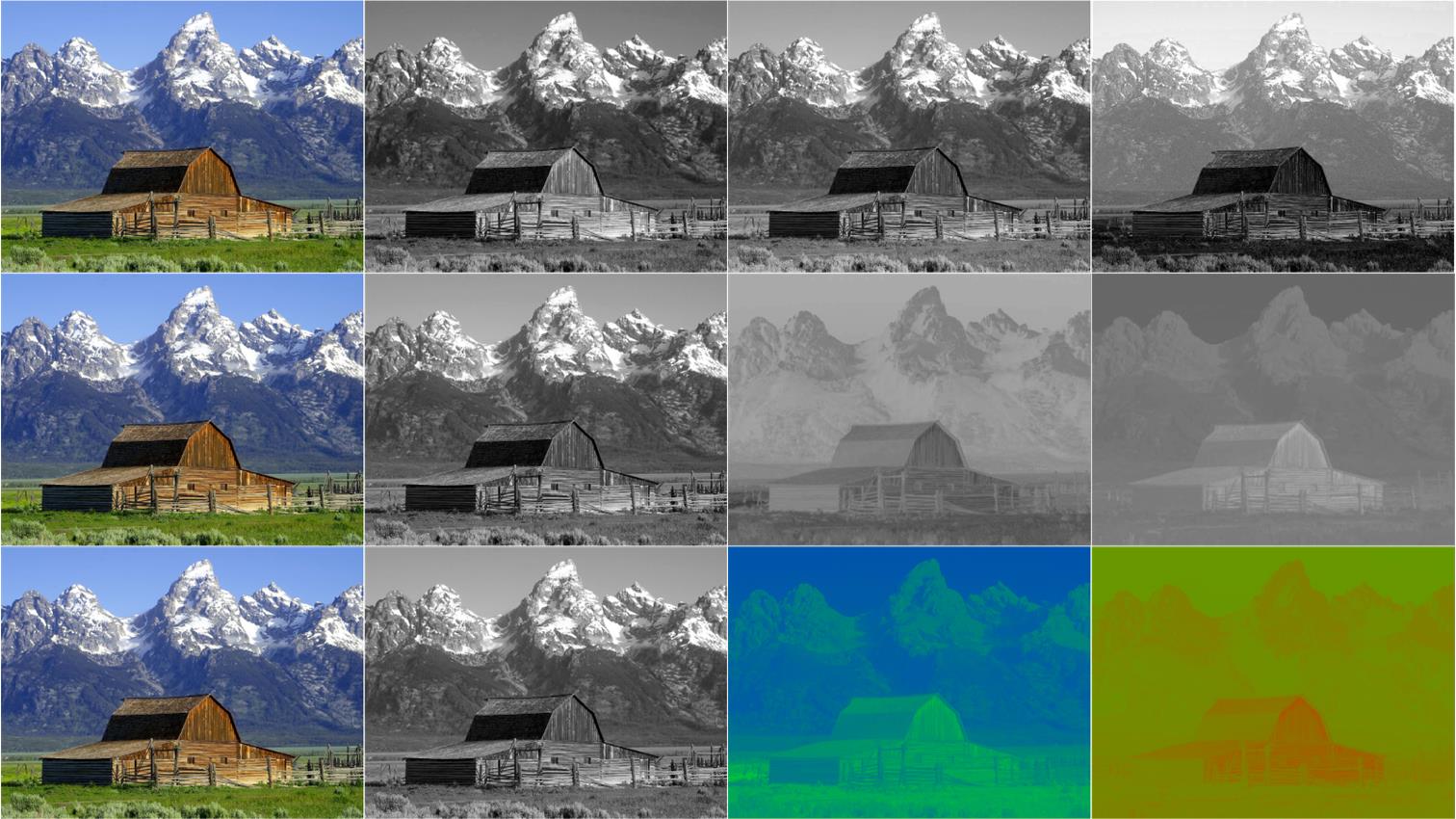}
  \caption{Three images in the first column are original RGB color images. Three images in the right of the first row are R channel, G channel and B channel of the original image respectively saved as gray images, three channels all constitutes the luminance information and color information. Three images in the right of the second row are Y channel, Cr channel and Cb channel respectively saved as gray images, and three images in the right of the third row are also Y channel, Cr channel and Cb channel respectively from Wikipedia. We can see that, the Y channel constitutes only the luminance information and semantic information, and the color information about chrominance and chroma are all in the Cr channel and the Cb channel.}
  \label{fig:rgb-yuv}
\end{figure}

Works of Baluja~\cite{nips} and Atique~\cite{Atique} have implemented the entire hiding and revealing procedure, while their stego images' color is distorted as shown in Fig.~\ref{fig:LFWshow}. To against this weakness, we select a new steganography position. As shown in Fig.~\ref{fig:rgb-yuv}, a color image in the RGB color space can be divided into R, G and B channels, and each channel contains both semantic information and color information. When converted to the YCrCb color space, a color image can be divided into Y, Cr and Cb channels. The Y channel only contains part of semantic information, luminance information and no color information, Cr and Cb channels contain part of semantic information and all color information. To guarantee no color distortion, we conceal the secret image only in the Y channel and all color information are saved into  the stego image. In addition, we select gray images as our secret images thus decreasing the secret information by $\frac{2}{3}$.  

When embedding, the color image is converted to the YCrCb color space, then the Y channel and the gray secret image are concatenated together and then are input to the encoder network. After hiding, the encoder's output and the cover image's CrCb channels constitute the color stego image. When revealing, we get the revealed secret image through decoding the Y channel of the stego image. Besides, the transformation between the RGB color space and the YCrCb color space is just the weighted computation of three channels and doesn't affect the backpropagation. So we can finish this tranformation during the entire hiding and revealing process. The encoder-decoder architecture can be trained end-to-end, which is called as the basic model. 

\subsection{Basic Model}\label{sec:3.2}

Conventional or classic image stegnography are usually designed in a heuristic way. Generally, these algorithms decide whether to conceal information into a pixel of the cover image and how to conceal 1 bit information into a pixel. So the key of the classic steganography methods is well hand-crafted algorithms, but all of these algorithms need lots of expertise and this is very difficult for us. The best solution is to mix the secret image with the cover image very well without too much expertise. Deep learning, represented by convolutional neural networks, is a good way to achieve this exactly. What we need to do is to design the structure of the encoder and the decoder as described below. 

\begin{figure}
  \centering
  \includegraphics[width=0.5\textwidth]{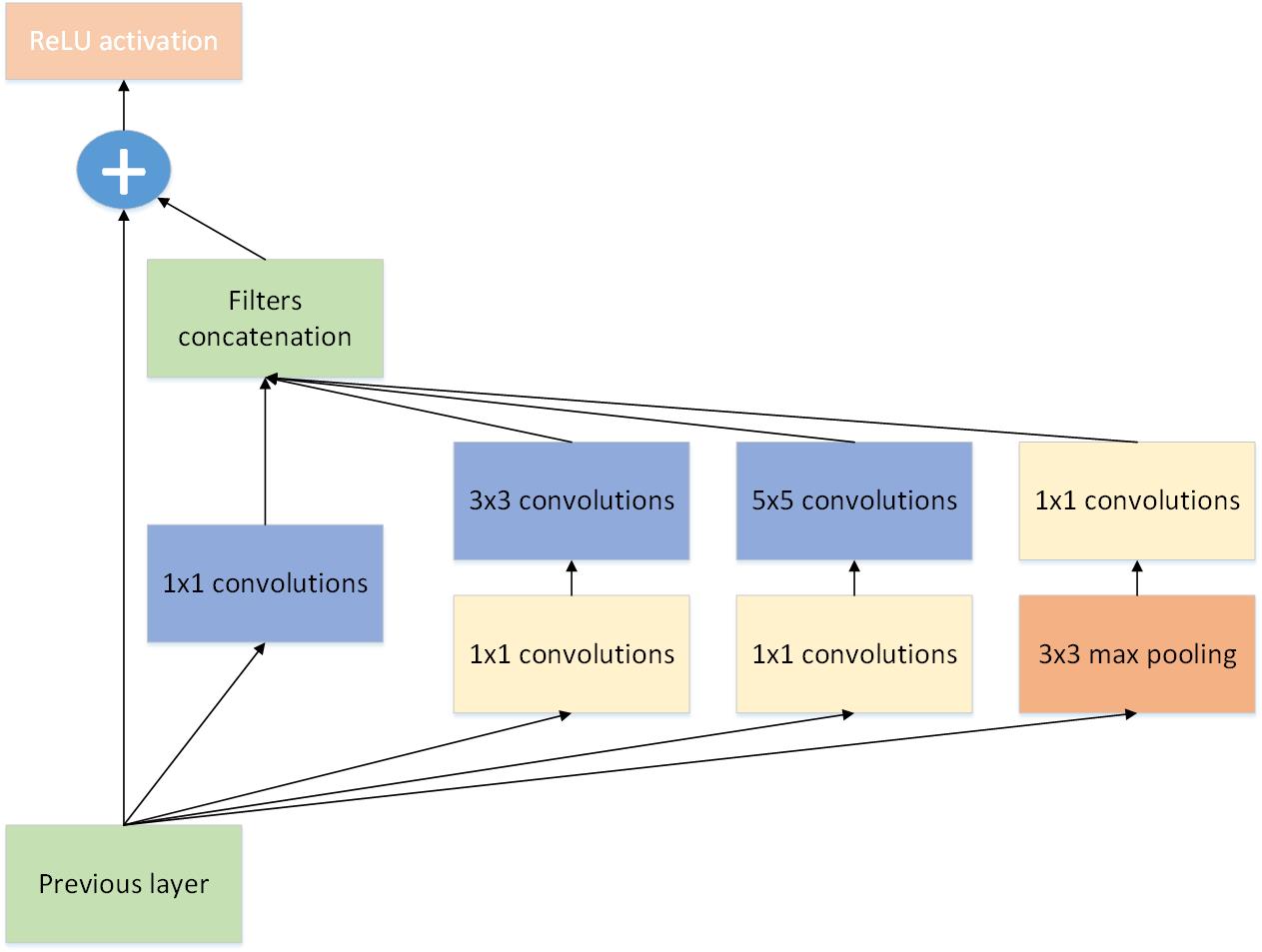}
\caption{The inception module with residual shortcut we use in our work.}
\label{fig:inception}
\end{figure}

\begin{table}
\caption{Architecture details of the encoder network: ConvBlock1 represents $3\times3$ Conv+BN+LeakyReLU, ConvBlock2 represents $1\times1$ Conv+Tanh, InceptionBlock represents the inception module with residual shortcut as shown in Fig.~\ref{fig:inception}.}
\label{tab:Encoder-Network}
\begin{tabular}{lcc}
\hline\noalign{\smallskip}
Layers & process & Output size \\
\noalign{\smallskip}\hline\noalign{\smallskip}
Input & /  & $2\times256\times256$ \\
Layer 1 & ConvBlock1  & $16\times256\times256$ \\
Layer 2 & InceptionBlock  &  $32\times256\times256$ \\
Layer 3 & InceptionBlock  &  $64\times256\times256$ \\
Layer 4 & InceptionBlock  &  $128\times256\times256$ \\
Layer 5 & InceptionBlock  &  $256\times256\times256$ \\
Layer 6 & InceptionBlock  &  $128\times256\times256$ \\
Layer 7 & InceptionBlock  &  $64\times256\times256$ \\
Layer 8 & InceptionBlock  &  $32\times256\times256$ \\
Layer 9 & ConvBlock1 & $16\times256\times256$ \\
Output & ConvBlock2 & $1\times256\times256$ \\
\noalign{\smallskip}\hline
\end{tabular}
\end{table}

Based on such a starting point, we introduce the inception module~\cite{inception} in our encoder network. The inception module has excellent performance on the ImageNet classification task, which contains several convolution kernels with different kernel sizes as shown in Fig.~\ref{fig:inception}. Such a model structure can fuse feature maps with different receptive field sizes very well. As shown in both residual networks~\cite{resnet} and batch normalization~\cite{BN}, a model with these modifications can achieve the performance with significantly fewer training steps comparing to its original version. So we introduce both residual module and batch normalization into the encoder network to speed up the training procedure. The detail structure of the encoder is described in Tab.~\ref{tab:Encoder-Network}. When using MSE as the metric on LFW dataset, we use our model to train for 30 epochs to get the performance Atique's model can achieve while training for 50 epochs.

\begin{table}
\caption{Architecture details of the decoder network: ConvBlock1 represents $3\times3$ Conv+BN+LeakyReLU, ConvBlock2 represents $1\times1$ Conv+Sigmoid.}
\label{tab:Decoder-Network}
\begin{tabular}{lcc}
\hline\noalign{\smallskip}
Layers & process & Output size \\
\noalign{\smallskip}\hline\noalign{\smallskip}
Input & /  & $1\times256\times256$ \\
Layer 1 & ConvBlock1  & $32\times256\times256$ \\
Layer 2 & ConvBlock1  &  $64\times256\times256$ \\
Layer 3 & ConvBlock1  &  $128\times256\times256$ \\
Layer 4 & ConvBlock1  &  $64\times256\times256$ \\
Layer 5 & ConvBlock1  &  $32\times256\times256$ \\
Output & ConvBlock2 & $1\times256\times256$ \\
\noalign{\smallskip}\hline
\end{tabular}
\end{table}

On the other hand, we need a structure to reveal the secret image out automatically. So we use a fully convolutional network as the decoder network. Feature maps output by each convolutional layer have the same size. To speed up training, we add a batch normalization layer after each convolutional layer other than the last layer.  Details of the decoder network are described in Tab.~\ref{tab:Decoder-Network}.

\subsection{Our steganalyzer}\label{sec:3.3}

Works of Baluja and Atique didn't consider the security problem, while the security is the keypoint in steganography. In our work, we want to take the steganalysis into account automatically throughout training the basic model. 

Denoting $\mathcal{C}$ as the set of all cover images $c$, the selection of cover images from $\mathcal{C}$ can be described by a random variable $c$ on $\mathcal{C}$ with probability distribution function (pdf) $P$. Assuming the cover images are selected with pdf $P$ and embedded with a secret image which is chosen from its corresponding set, the set of all stego images is again a random variable $s$ on $\mathcal{C}$ with pdf $Q$. The statistical detectability can be measured by the Kullback-Leibler divergence~\cite{KL} shown in \eqref{kl} or the Jensen-Shannon divergence shown in \eqref{js}.

\begin{equation}
  KL(P||Q) = \sum_{c\in \mathcal{C}} P(c) log\frac{P(c)}{Q(c)} \label{kl}
\end{equation}

\begin{equation}
  JS(P||Q) = \frac{1}{2}KL(P\|\frac{P+Q}{2}) + \frac{1}{2}KL(Q\|\frac{P+Q}{2}) \label{js}
\end{equation}

The KL divergence or the JS divergence is a very fundamental quantity because it provides bounds on the best possible steganalyzer one can build ~\cite{information theory}. So the keypoint for us is how to decrease the divergence. The generative adversarial networks (GAN) are well-designed in theory to achieve this exactly. The objective of the original GAN is to minimize the JS divergence \eqref{js}, a variant of the GAN is to minimize the KL divergence \eqref{kl}. The generator network $G$, which input is a noise $z$, tries to transform the input to a data sample which is similar to the real sample. The discriminator network $D$, which input is the real data or the fake data generated by the generator network, determines the difference between the real and fake samples. $D$ and $G$ play a two-player minmax game with the value function \eqref{V1}.

\begin{equation}
  \min_{G}\max_{D}=E_{x\sim p_{data}(x)}[logD(x)] + E_{z\sim p_z(z)}[log(1-D(G(z)))] \label{V1}
\end{equation}

Now we introduce the generative adversarial networks into our architecture. The basic model can finish the entire hiding and revealing process, so we use the basic model as the generator, and introduce a CNN-based steganalysis model as the discriminator and the steganalyzer. So the value function in our work becomes \eqref{V2}, where $D$ represents the steganalyzer network, $G$ represents the basic model, $x$, $s$ and $G(x, s)$ represent the cover image, the secret image and the generated stego image respectively.

\begin{equation}
  \min_{G}\max_{D}=E_{x\sim P(x)}[logD(x)] + E_{x\sim P(x), s\sim P(s)}[log(1-D(G(x, s)))] \label{V2}
\end{equation}

\begin{table}
\caption{Architecture details of the steganalyzer network: ConvBlock1 represents $3\times3$ Conv+BN+LeakyReLU+AvgPool, ConvBlock2 represents $1\times1$ Conv+BN and ConvBlock3 represents $1\times1$ Conv+BN+LeakyReLU. SPPBlock contains a SPP modeule and the FC represents a fully connected layer.}
\label{tab:Steganalyzer-Network}
\begin{tabular}{lcc}
\hline\noalign{\smallskip}
Layers & process & Output size \\
\noalign{\smallskip}\hline\noalign{\smallskip}
Input & /  & $3\times256\times256$ \\
Layer 1 & ConvBlock1  & $8\times128\times128$ \\
Layer 2 & ConvBlock1  &  $16\times64\times64$ \\
Layer 3 & ConvBlock2  &  $32\times32\times32$ \\
Layer 4 & ConvBlock2  &  $64\times16\times16$ \\
Layer 5 & ConvBlock3  &  $128\times8\times8$ \\
Layer 6 & SPPBlock  &  $2688\times1$ \\
Layer 7 & FC  &  $128\times1$ \\
Layer 8 & FC  &  $2\times1$ \\
\noalign{\smallskip}\hline
\end{tabular}
\end{table}

Xu et al.\cite{XuNet} studied the design of CNN structure specific for image steganalysis applications and proposed XuNet. XuNet embeds an absolute activation (ABS) in the first convolutional layer to improve the statistical modeling, applies the TanH activation function in early stages of networks to prevent overfitting, and adds batch normalization (BN) before each nonlinear activation layer. This well-designed CNN provides excellent detection performance in steganalysis. So we design our steganalyzer based on XuNet and adapt it to fit our stego images. In addition, we use the spatial pyramid pooling (SPP) module to replace the global average pooling layer. The spatial pyramid pooling (SPP) module\cite{spp} and its variants, which play a huge role in models for objection detection and semantic segmentation, break the limit of fully connected layers, so that images with arbitrary sizes can be input into convolutional networks with fully connected layers. On the other hand, the SPP module can extract more features from different receptive fields, thus improving the performance. Our steganalyzer's detail architecture is shown in Tab.~\ref{tab:Steganalyzer-Network}.

\subsection{Mixed Loss Function}\label{sec:3.4}

In previous works, Baluja~\cite{nips} used the mean square error (\mse) between the pixels of original images and the pixels of reconstructed images as the metric \eqref{eq1}. Where c and s are the cover and secret images respectively, $c^{'}$ and $s^{'}$ are the stego and revealed secret images respectively, and $\beta$ is how to weight their reconstruction errors. In particular, we should note that the error term $||c-c'||$ doesn't apply to the weights of the decoder network. On the other hand, both the encoder network and the decoder network receive the error signal $\beta||s-s'||$ for reconstructing the secret image. 

\begin{equation}
  L(c, c', s, s')=\| c - c'\| + \beta \| s - s'\|\label{eq1}
\end{equation}

However, the \mse just penalizes large error of two images' corresponding pixels but disregards the underlying structure in images. The human visual system (HVS) is more sensitive to luminance and color variations in texture-less regions. Zhao et al.~\cite{Zhao} analyzed the importance of perceptually-motivated losses when the resulting image of image restoration tasks is evaluated by a human observer. They compared the performance of several losses and proposed a novel, differentiable error function. Inspired by their work, we introduce the structure similarity index (\ssim)~\cite{SSIM} and its variant, the multi-scale structure similarity index (\mssim)~\cite{MS-SSIM} into our metric. 

The \ssim index separates the task of similarity measurement into three comparisons: luminance, contrast and structure. The luminance, contrast and structure similarity of two images are measured by \eqref{eq2}, \eqref{eq3} and \eqref{eq4} respectively. Where $\mu_x$ and $\mu_y$ are pixel average of image x and image y, $\theta_x$ and $\theta_y$ are pixel deviation of image x and image y, and $\theta_{xy}$ is the standard variance of image x and y. In addition,  $C_1$, $C_2$ and $C_3$ are constants included to avoid instability when denominators are close to zero. The total calculation method of \ssim is shown in \eqref{eq5}, where $l>0, m>0, n>0$ are parameters used to adjust the relative importance of three components. More detail introduction to \ssim can be found in ~\cite{SSIM}. The value range of the \ssim index is $[0, 1]$. The higher the index is, the more similar the two images are. So we use $1 - SSIM(x, y)$ in our loss function to measure the difference of two images. And the \mssim~\cite{MS-SSIM} is an enhanced variant of the \ssim index, so we also introduce it into our loss function (We use $MSSIM$ in functions to represent MS-SSIM). 

\begin{equation}
  L(x, y) = \frac{2\mu_x\mu_y+C_1}{\mu_x^2+\mu_y^2+C_1}\label{eq2}
\end{equation}

\begin{equation}
  C(x, y) = \frac{2\theta_x\theta_y+C_2}{\theta_x^2+\theta_y^2+C_2}\label{eq3}
\end{equation}

\begin{equation}
  S(x, y) = \frac{\theta_{xy}+C_3}{\theta_x\theta_y + C_3}\label{eq4}
\end{equation}

\begin{equation}
  SSIM(x, y) = [L(x, y)]^{l} \cdot [C(x, y)]^{m} \cdot [S(x, y)]^{n}\label{eq5}
\end{equation}

Considering pixel value differences and structure differences simultaneously, we put \mse, \ssim and \mssim together. So, the metric for the basic steganography network in our framework is as below: 

\begin{equation}
\begin{aligned}
  L(c, c') = \alpha (1 - SSIM(c, c')) \\  
  + (1 - \alpha)(1 - MSSIM(c, c')) \\ 
  + \beta MSE(c, c')\label{eq6}
\end{aligned}
\end{equation}

\begin{equation}
\begin{aligned}
  L(s, s') = \alpha (1 - SSIM(s, s')) \\ 
  + (1 - \alpha)(1 - MSSIM(s, s')) \\ 
  + \beta MSE(s, s')\label{eq7}
\end{aligned}
\end{equation}

\begin{equation}
  L(c, c', s, s') = L(c, c') + \gamma L(s, s')\label{eq8}
\end{equation}

Where $\alpha$ and $\beta$ are hyperparameters to weigh influences of three metrics and $\gamma$ is a hyperparameter to trade off the quality of stego images and revealed secret images. Experiment results in \secref{4} will compare the performance of different loss functions. 

\section{Experiments and Results}\label{sec:4}

In this section, we'll introduce our experiment details and results. Firstly, the datasets we used are LFW~\cite{lfw}, Pascal VOC 2012~\cite{voc} and ImageNet~\cite{imagenet}. The Labeled Faces in the Wild (LFW) contains more than 13000 face images belonging to 1680 people collected from the web. 10k images were selected from LFW and constituted 5k cover-secret image pairs as our training set, others of LFW were as our validation set. Pascal VOC 2012 is a dataset designed for object detection and semantic segmentation, we selected 16k images randomly to constitute 8k cover-secret image pairs as our training set and selected 5k images from the remaining part as our validation set. To further verify our model's performance on the big dataset, we did similar experiments on a subset of the ImageNet. Limited by the computing power, we only used the validation set of ImageNet as our training set which contains 50k images, these images constituted 25k cover-secret image pairs randomly. Then we selected 30k images from the test set of ImageNet as our validation set. 

We used \ssim~\cite{SSIM}, Peak Signal to Noise Ration (\psnr) as metrics to measure our model's performance. It is widely accepted that the \psnr doesn't correlate well with the human's perception of image quality~\cite{Zhang}, so we just used it as a reference. In addition, we designed a CNN-based steganalyzer specially to measure our model's security. 

All settings of our model on three datasets were the same. All parameters of our model were initialized by the Xavier initialization~\cite{xavier} and the initial learning rate was set as 1e-4 and was descended during training after 20 epochs. The batch size was set as 4 limited by the computing power, and we used Adam to optimize our basic model. After several attempts, we set $\alpha$, $\beta$ and $\gamma$ of the loss function as 0.5, 0.3 and 0.85 respectively, which can trade off the quality of stego images and revealed secret images very well. Because our secret message is an image, so we don't need to reveal out the secret image completely. Certainly, you can set $\gamma$ higher if you want better revealed secret images. The size of all images we used is $256 \times 256$, and the capacity of our model is 8\bpp (it is equivalent to that we hide a pixel (8 bits) in a pixel). 

\begin{table}
\centering
\caption{We use several loss functions to train our basic model on LFW for 50 epochs. $\mse+\ssim$ represents a mix loss of \mse and \ssim, others are similar, and $revealed$ represents revealed secret images. We can see that the mix loss of \mse, \ssim and \mssim is superior than others.}
\label{tab:loss-compare}
\resizebox{0.5\textwidth}{!}{    
\begin{tabular}{lllllll}
\hline\noalign{\smallskip}
Loss Function & \begin{tabular}[c]{@{}l@{}}Stego-Cover\\ \psnr (db)\end{tabular} & \begin{tabular}[c]{@{}l@{}}Revealed-Secret\\ \psnr (db)\end{tabular} & \begin{tabular}[c]{@{}l@{}}Stego-Cover\\ \ssim \end{tabular} & \begin{tabular}[c]{@{}l@{}}Revealed-Secret\\ \ssim \end{tabular} \\ 
\noalign{\smallskip}\hline\noalign{\smallskip}
\mse & 27.97 & 26.30 & 0.8592 & 0.8391 \\
\ssim & 21.71 & 22.76 & 0.8877 & 0.8466 \\
\mse+\ssim & 27.12 & 26.71 & 0.8921 & 0.8805 \\
\mse+\mssim & 23.92 & 25.97 & 0.8287 & 0.8832 \\
\mse+\ssim+\mssim & 26.72 & 25.97 & \textbf{0.9305} & \textbf{0.9160} \\
\noalign{\smallskip}\hline
\end{tabular}}
\end{table}

\begin{table}
\centering
\caption{We can see that the \ssim index between stego images and their corresponding cover images of \isgan is higher than our basic model and Atique's work\cite{Atique}.}
\label{tab:dataresults}
\resizebox{0.5\textwidth}{!}{      
\begin{tabular}{lllllllll}
\hline\noalign{\smallskip}
Model & \begin{tabular}[c]{@{}l@{}}Cover\\ Image\end{tabular} & \begin{tabular}[c]{@{}l@{}}Secret\\ Image\end{tabular} & \begin{tabular}[c]{@{}l@{}}Stego-Cover\\ \psnr (db)\end{tabular} & \begin{tabular}[c]{@{}l@{}}Revealed-Secret\\ \psnr (db)\end{tabular} & \begin{tabular}[c]{@{}l@{}}Stego-Cover\\ \ssim \end{tabular} & \begin{tabular}[c]{@{}l@{}}Revealed-Secret\\ \ssim \end{tabular} \\
\noalign{\smallskip}\hline\noalign{\smallskip}
Atique's model & \lfw & \lfw & 33.7 & 39.9 & 0.95 \\
Basic model & \lfw & \lfw & \textbf{34.28} & 33.53 & \textbf{0.9529} & 0.9453 \\
ISGAN & \lfw & \lfw & \textbf{34.63} & 33.63 & \textbf{0.9573} & 0.9429 \\
Atique's model & \imagenet & \imagenet & 32.9 & 36.6 & 0.96 & \textbf{0.96} \\
Basic model & \imagenet & \imagenet & \textbf{34.57} & 33.53 & \textbf{0.9634} & 0.9510 \\
ISGAN & \imagenet & \imagenet & \textbf{34.89} & 33.42 & \textbf{0.9681} & 0.9474 \\
Atique's model & \pascal & \pascal & 33.7 & 35.9 & 0.96 & \textbf{0.95} \\
Basic model & \pascal & \pascal & \textbf{33.79} & 33.47 & \textbf{0.9617} & 0.9475 \\
ISGAN & \pascal & \pascal & \textbf{34.49} & 33.31 & \textbf{0.9661} & 0.9467 \\
Atique's model & \pascal & \lfw & 33.8 & 37.7 & 0.96 & 0.95 \\
Basic model & \pascal & \lfw & \textbf{33.85} & 37.68 & \textbf{0.9612} & 0.9503 \\
ISGAN & \pascal & \lfw & \textbf{34.45} & 37.59 & \textbf{0.9647} & 0.9495 \\
ISGAN & \imagenet & \pascal & \textbf{34.57} & 36.58 & \textbf{0.9652} & 0.9495 \\
\noalign{\smallskip}\hline
\end{tabular}}
\end{table}

\begin{figure}
\centering
   \includegraphics[width=\linewidth]{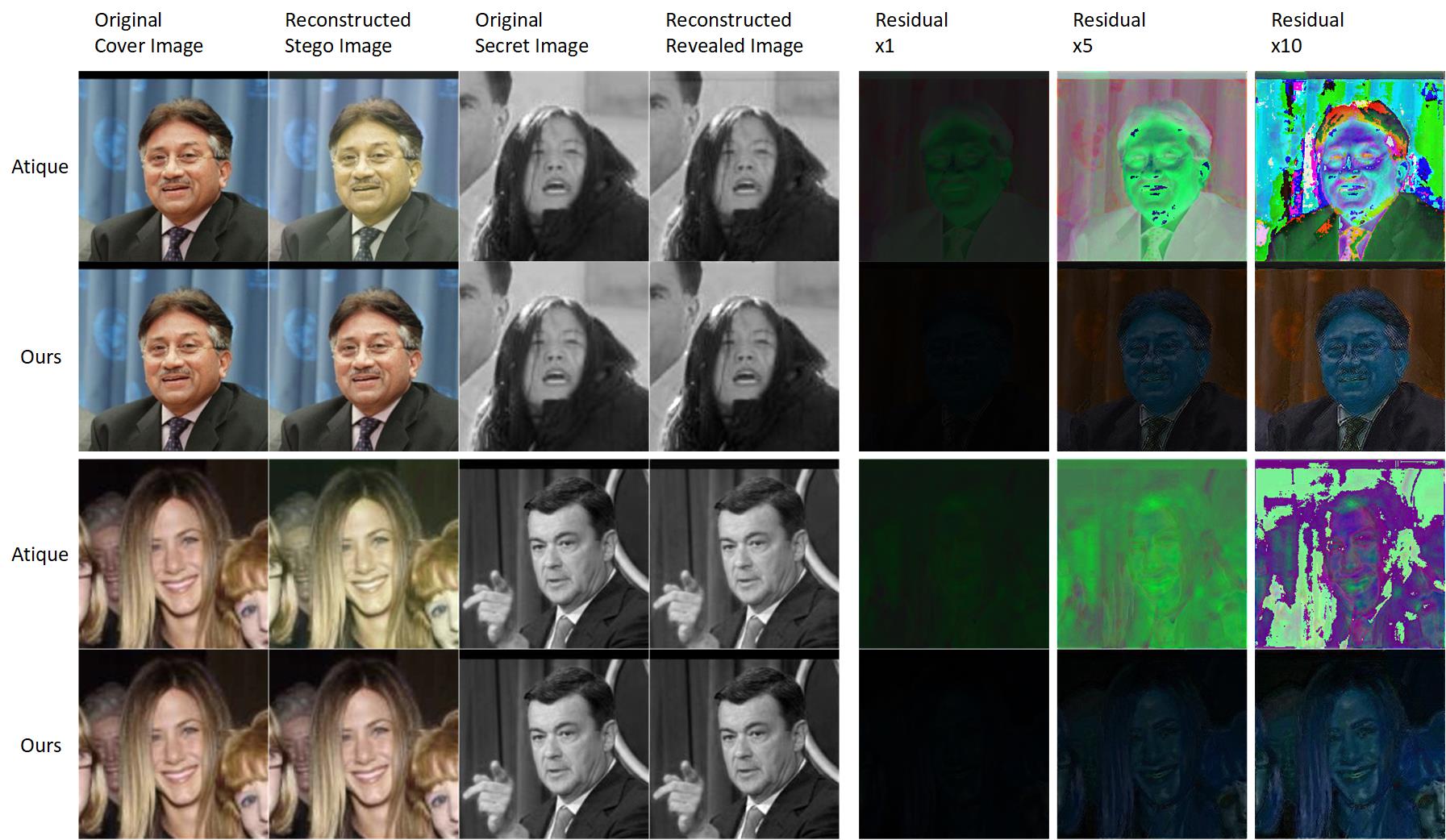}
   \caption{Two examples on LFW. We can see that our stego images are almost same as cover images, while Atique's stego images are yellowing. By analyzing residuals between stego images and cover images, we can see that our stego images are more similar to cover images than Atique's results.}
\label{fig:LFWshow}
\end{figure}

\begin{figure}
\centering
   \includegraphics[width=\linewidth]{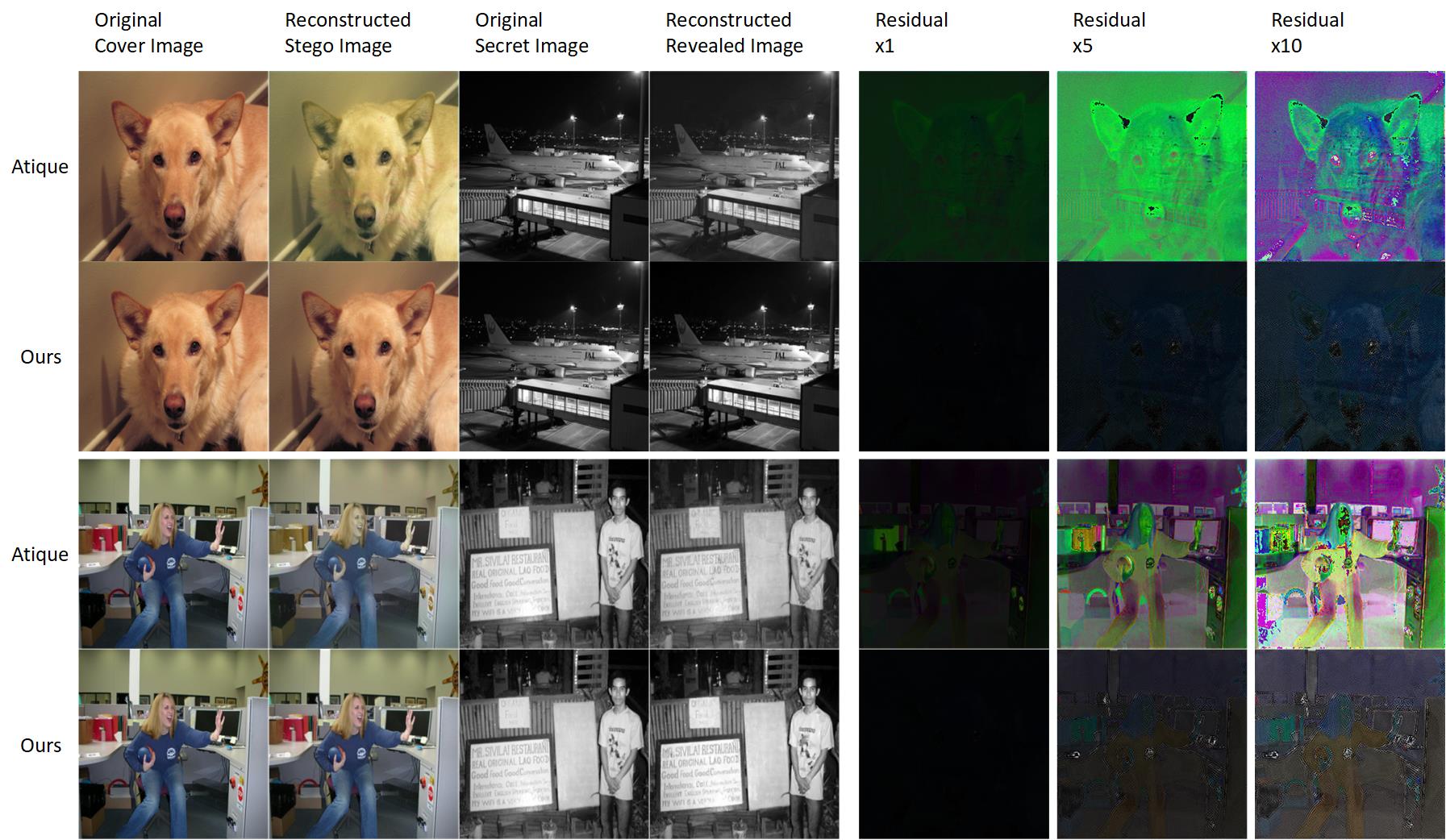}
   \caption{Two examples on Pascal VOC12. We can see that our stego images are almost same as cover images, while Atique's stego images are yellowing. By analyzing residuals between stego images and cover images, we can even distinguish the outline of secret images from Atique's residual images, while our residual images are blurrier.}
\label{fig:VOCshow}
\end{figure}

\begin{figure}
\centering
   \includegraphics[width=\linewidth]{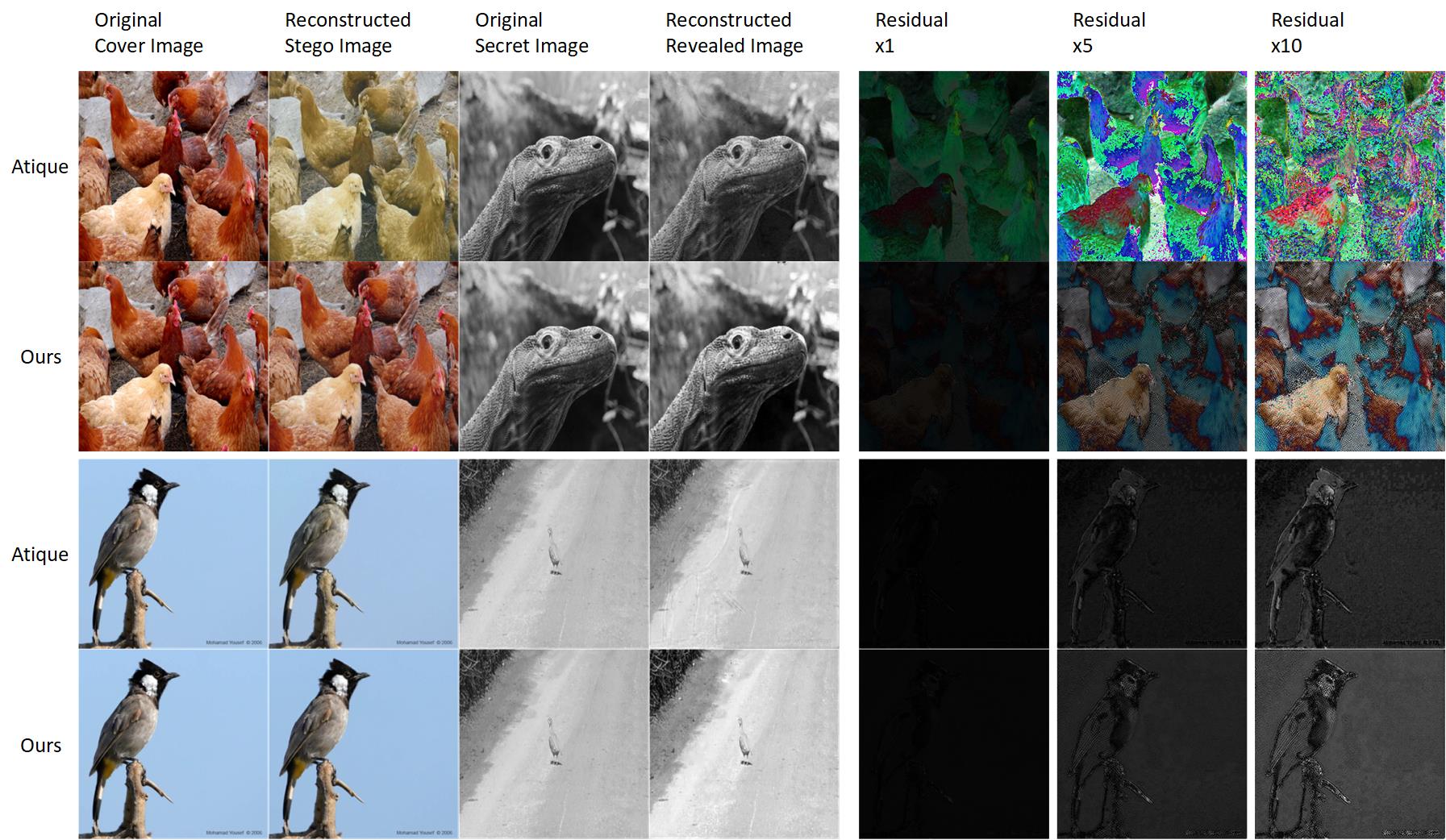}
   \caption{Two examples on ImageNet. We can see that our stego images are almost same as cover images, while Atique's stego images are yellowing. Residual images between stego images and cover images show that our stego images are more similar to cover images than Atique's results.}
\label{fig:ImageNetshow}
\end{figure}

As shown in Tab.~\ref{tab:loss-compare}, we do several experiments with different loss functions on the LFW, the result demonstrates that our proposed mixed loss function is superior to others. Tab.~\ref{tab:dataresults} describes final results of our model on three datasets, we can see that the invisibility of our model get a little improvement, while our model's performance is superior to Atique's work intuitively as shown in Fig.~\ref{fig:LFWshow}, ~\ref{fig:VOCshow} and ~\ref{fig:ImageNetshow}. Stego images generated by our model are complete similar to corresponding cover images in semantic and color, this is not reflected by \ssim. On the training set, the average \ssim index between stego images generated by our model and their corresponding cover images is more than 0.985, and the average \ssim index between revealed images and their corresponding secret images is more than 0.97. In practice, we can use several cover images to conceal one secret image and choose the best stego image to transfer on the Internet. 

On the other hand, by analyzing the detail difference between cover images and stego images, we can see that our residual images are darker than Atique’s, which means that our stego images are more similar to cover images and \isgan has stronger invisibility. Additionally, from Atique’s residual images we can even distinguish secret images’ outline, while our residual images are blurrier. So these residual images can also prove that our \isgan is securer.

When training \isgan, we referred some tricks from previous works~\cite{tricks}. We flipped labels when training our basic model, replaced the ReLU activation function by the LeakyReLU function, optimized the generator by Adam, optimized the steganalyzer by SGD and applied the L2 normalization to inhibit overfitting. These tricks helped us to speed up training and get better results.

To prove the improvement of the security produced by generative adversarial networks, we designed a new experiment. We used a well-trained basic model to generate 5000 stego images on LFW. These 5000 stego images and their corresponding cover images constituted a tiny dataset. We designed a new CNN-based model as a binary classifier to train on the tiny dataset. After training, we used this model to recognize stego images out from another tiny dataset which contains 2000 stego images generated by \isgan and their corresponding cover images. Similar experiments were done on the other two datasets. The results can be seen from Tab.~\ref{tab:security-results}. \isgan strengthens indeed the security of our basic model. And with the training going, we can see that the security of \isgan is improving slowly. 

\begin{table}[tbp]
  \begin{center}
  \resizebox{0.5\textwidth}{!}{
\begin{tabular}{lllll}
\hline\noalign{\smallskip}
Dataset &  Basic Model & ISGAN (50) & ISGAN (100) & ISGAN (150) \\
\noalign{\smallskip}\hline\noalign{\smallskip}
LFW & 0.8305 & 0.8059 & 0.7887 & \textbf{0.7825} \\
Pascal-VOC12 & 0.7953 & 0.769 & 0.756 & \textbf{0.7438} \\
ImageNet & 0.7814 & 0.7655 & 0.7462   & \textbf{0.7360} \\
\noalign{\smallskip}\hline
\end{tabular}
}
  \end{center}
  \caption{Accuracy of CNN-based steganalysis model on tiny-datasets generated by basic model and \isgan training for different epochs. Along with the training going, we can see that the security of \isgan is improving slowly.}
\label{tab:security-results}
\end{table}


\section{Discussion and Conclusion}\label{sec:5}

\begin{figure}
\centering
   \includegraphics[width=1.0\linewidth]{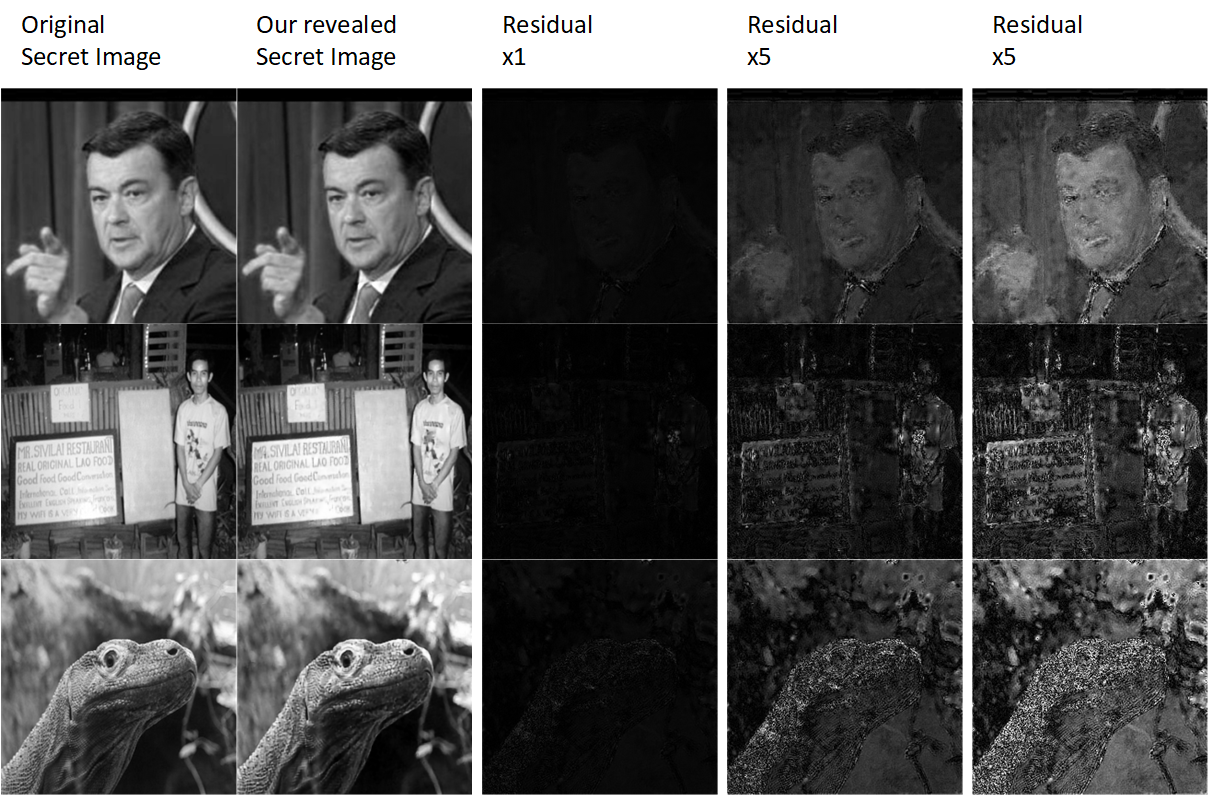}
   \caption{Secret images' residual image on three datasets. We can see that there are differences between original secret images and our revealed secret images, which means that \isgan is a lossy steganography.}
   \label{fig:secret-residual}
\end{figure}

Fig.~\ref{fig:secret-residual} shows the difference between revealed images and their corresponding secret images. It shows that this kind of model cannot reveal out secret images completely. This is accepted as the information in the secret image is very redundant. However, it is unsuitable for tasks which need to reveal the secret information out completely.

As we described before, \isgan can conceal a gray secret image into a color cover image with the same size excellently and generate stego images which are almost the same as cover images in semantic and color. By means of the adversarial training, the security is improved. In addition, experiment results demonstrate that our mixed loss function based on SSIM can achieve the state-of-art performance on the steganography task. 

In addition, our steganography is done in the spatial domain and stego images must be lossless, otherwise some parts of the secret image will be lost. There may be methods to address this problem. It doesn't matter if the stego image is sightly lossy since the secret image is inherently redundant. Some noise can be added into the stego images to simulate the image loss caused by the transmission during training. Then our decoder network should be modified to fit both the revealing process and the image enhancement process together. In our future work, we'll try to figure out this problem and improve our model's robustness.

\end{document}